\documentclass{aastex63}

\usepackage{amsmath}
\usepackage{xcolor}

\begin{document}

\title{Stability and Convergence of Nuclear Detonations in White Dwarf Collisions}

\author{Peter Anninos}
\affil{Lawrence Livermore National Laboratory, Livermore, CA 94550, USA}

\author{David Cruz-Lopez}
\affil{Department of Astronomy, University of Illinois, Urbana, IL 61801, USA}

\author{Brighten Jiang}
\affil{Department of Astronomy, University of Florida, Gainesville, FL 32611, USA}

\author{Emanuel Gordis}
\affil{Department of Physics, Cornell University, Ithaca, NY 14853, USA}

\begin{abstract}
We investigate the numerical stability of thermonuclear detonations in 
1D accelerated reactive shocks and 2D binary collisions of equal mass,
magnetized and unmagnetized white dwarf stars. To achieve high resolution at initiation sites, we devised
geometric gridding and mesh velocity strategies specially adapted to the unique
requirements of head-on collisional geometries, scenarios in which one expects maximum
production of iron-group products. We study effects of grid resolution and the
limiting of temperature, energy generation, and reactants for different stellar masses, separations,
magnetic fields, initial compositions, detonation mechanisms, 
and limiter parameters across a range of cell sizes from 1 to 100 km.
Our results set bounds on the parameter space of limiter amplitudes for which both temperature
and energy limiting procedures yield consistent and monotonically convergent solutions.
Within these bounds we find grid resolutions of 5 km or better are necessary for uncertainties 
in total released energy and iron-group products to drop below 10\%. Intermediate mass products
(e.g., calcium) exhibit similar convergence trends but with somewhat greater uncertainty.
These conclusions apply equally to pure C/O WDs, multi-species compositions (including helium shells),
magnetized and unmagnetized cores, and either single or multiple detonation scenarios.
\end{abstract}


\section{Introduction}
\label{sec:intro}

Although there is no universally accepted consensus on the thermonuclear initiation mechanism
in Type Ia supernovae, it is generally believed explosions result from the nuclear detonation
of white dwarf (WD) stars, even if the mechanism itself is not fully understood.
For the case of colliding WDs, however, initiation is well-known to be triggered by shocks.
\replaced{
What perhaps remains uncertain in these scenarios is the degree to which spatial (and temporal) resolution
affects ignition, burn product distributions, and subsequently observed luminosities.
} {
What perhaps remains uncertain in these scenarios is the degree to which spatial (and temporal) resolution
affects ignition and burn product distributions, or even whether ignition is initiated
promptly at first contact, potentially affecting luminosities and their similarity
to Ia-like events.
}
This is an especially important consideration for multi-dimensional simulations straining
computational resources to
simultaneously cover the small detonation scales together with global binary dynamics,
separated easily by more than four orders of magnitude.

In a recent study, \citet{Katz19} (henceforth referred to as KZ19) proposed a set of one-dimensional shock tube problems,
replicating conditions from their multi-dimensional work on WD collisions, as a proxy test to assess convergent behavior.
Based on results from that study, they caution resolutions finer than a kilometer may be required
to achieve convergent and consistent results.
\citet{Kushnir20a} (KK20) however contradict this conclusion and instead
claim convergence for the same test problems at coarser resolutions of a few kilometers.
In our own proxy studies of tidal disruption events (TDEs) we found better than 10\% convergence
for some attributes (like nickel production) below 10 km resolution, 
but significantly greater uncertainty in many transient products even at resolutions as low as
1 km \citep{Anninos22}. 
\deleted{
Apart from transients, our results (for energy release and nickel production) are generally
are generally consistent with KK20 despite the different interactions and ignition scenarios.
}

Summarizing the literature survey by KZ19, we note most
calculations of WD collisions were run at resolutions between 10-100 km with 
SPH codes \citep{Rosswog09,Raskin10,Loren-Aguilar10,Garcia13},
and 50-500 km with grid-based codes \citep{Rosswog09,Hawley12}
with notable exceptions for calculations performed
in axisymmetry \citep{Kushnir13,Papish16}. All of these studies managed to convert large amounts of material
to iron-group elements, but did not agree quantitatively on final outcomes.
In some cases the consistency (or convergence) of behaviors with resolution was not discussed. In others
claims were made suggesting convergent behavior but either over a small dynamical range or over
a larger respectable range that exhibited converging trends but nonetheless failed to 
fully plateau important diagnostics like nickel production.

\citet{Kushnir20a} argue many of these studies suffered premature ignition from numerically unstable evolutions
attributed to zone sizes (or smoothing lengths) being substantially larger than scales over which detonations develop,
a common and well-known deficiency with global modeling \citep{Niemeyer97,Ropke07,Seitenzahl09}.
To put this scale disparity into perspective, hotspot sizes vary widely based on
local physical conditions and compositions \citep{Seitenzahl09,Holcomb13,Garg17},
but can be less than a kilometer at conditions observed in many of our calculations.
This is significantly smaller, often by orders of magnitude, than afforded resolutions.
KK20 suggest the enforcement of burn limiters stabilizes solutions consistent with convergent behavior, 
and the lack of such limiters might explain previously discrepant results.
\replaced{
Our own implementation of nucleosynthesis features a combination of temperature, energy deposition, and reactant limiters
as options utilized regularly in our TDE studies,
so we apply and explore their use in this different context along with a tailored gridding strategy
designed to concentrate computational resources near initiation sites.
} {
Our own implementation of nucleosynthesis features a combination of temperature, energy deposition, and reactant limiters
proven effective at stabilizing hydrodynamic interactions, and apply them to a tailored gridding strategy
designed to concentrate computational resources near initiation sites.
}

In addition to the one-dimensional shock tube tests proposed by \citet{Katz19}, we carry out 
systematic convergence studies of WD collisions in 2D cylindrical symmetry, including
effects of magnetic fields and helium shells triggering double detonations.
Due to conformally strict meshing requirements, all of our calculations are restricted to head-on equal mass collisions.
But we consider a couple of different stellar masses (0.6 and 1 $M\odot$),
a broad range of limiter parameters, toroidal and poloidal magnetic field configurations,
several initial compositions (pure C/O, Mesa-generated models, helium shells),
and importantly two decades of spatial resolution (1 - 100 km).

Section \ref{sec:methods} begins with a brief discussion of our numerical methods, physical models
(equation of state, reactive networks, initial data, etc.), and gridding strategies tuned to achieve high
spatial resolution at detonation sites.
Our results follow in section \ref{sec:results}, and we conclude with a brief summary
in section \ref{sec:conclusions}.

\section{Methods and Models}
\label{sec:methods}

All calculations are performed with the {\sc Cosmos++} code
\citep{Anninos05,Anninos17,Anninos20,Roth22}, which solves
the equations for self-gravitating, Newtonian or general relativistic radiation magneto-hydrodynamics coupled with
thermonuclear reactions and energy generation on unstructured, moving and adaptively refined (AMR and/or ALE) meshes.  
Our investigations of TDEs \citep{Anninos18,Anninos19,Anninos22} modeled binary encounters between WD stars and
black holes with general relativistic hydrodynamics. Relativistic effects are not important in this work,
so we solve instead the simpler Newtonian MHD equations together with Poisson's equation for self-gravity.

Thermodynamics is treated with a Helmholtz equation of state, accounting for radiation,
electron degeneracy, Coulomb corrections, relativistic effects, and electron-positron contributions,
It is based on the Torch model \citep{Timmes00a}, accommodating
arbitrary isotopic compositions, nuclear reaction networks, densities
spanning $10^{-12} \le \rho \le 10^{15}$ gm cm$^{-3}$,
and temperatures $10^{3} \le T \le 10^{13}$ K.

{\sc Cosmos++} supports several nuclear network options, including 7-, 19-, and 21-isotope alpha-chain
models \citep{Weaver78,Timmes99,Anninos18}, fully coupled with
hydrodynamics, advection of reactants, photodisintegration, electron capture, and nuclear energy coupling,
along with an assortment of choices for the integration of stiff differential equations.
We use the 19-isotope network throughout this work, which consists of the following nuclei:
$^{4}$He, $^{12}$C, $^{16}$O, $^{20}$Ne, $^{24}$Mg, $^{28}$Si, $^{32}$S, $^{36}$Ar, $^{40}$Ca, $^{44}$Ti,
$^{48}$Cr, $^{52}$Fe, and $^{56}$Ni, plus additional species ($^{1}$H, $^{3}$He, $^{14}$N, $^{54}$Fe) to
accommodate hydrogen burning, as well as photo-disintegration neutrons and protons.

The network is solved using a 4th order fully implicit method, incorporating a number of options
to limit burn energy generation and deposition $\delta e_{nuc}$ by either enforcing the energy condition
$|\delta e_{nuc}|/e \le F(\Delta t_{gas}, \Delta t_{cs}) = f_E ~[\Delta t_{gas}/c_{FL} \Delta t_{cs}]$ in each
cell every cycle, by constraining temperature variations $|\delta T|/T \le f_T$, 
by limiting fractional changes to molar abundances $|\delta Y_i|/Y_i \le f_Y$, or all three in any combination.
Here $\Delta t_{gas}$ and $\Delta t_{cs}$ are the hydrodynamic timestep and local sound crossing times
respectively, $c_{FL}$ is the Courant factor, 
$e$ is the gas internal energy density, and ($f_E$, $f_T$, $f_Y$) are constants less than unity.
\added{
These limiting options are applied across every hydrodynamic cycle, which typically evolves with
a much greater timestep than the characteristic time required for the nuclear network to remain
stable (without time-implicit solvers), or to deposit released energy
into cells at a rate satisfying hydrodynamic stability conditions (for example a sound
crossing time). Stability between hydrodynamics and nuclear energy production can be acheived
by either increasing grid resolution, or evolving with a small enough timestep preventing
excessive energy production. Although {\sc Cosmos++} offers an option to constrain
timesteps based on local estimates of energy production rates, it is not particularly
effective in practice and leads to excessively long runtimes, especially for multi-dimensional calculations.
Alternatively, burn limiters provide the necessary stability at hydrodynamic timescales
while maintaining desired monotonic convergence with resolution. They operate in a manner similar
to hydrodynamic timestep limiting, except the limit is applied to the network timestep which
has the effect of scaling reaction rates to prevent excessive energy generation.
Integration over these smaller reaction limiting timesteps is performed iteratively with
a bisection algorithm that converges on the burn timestep required to satisfy
energy production (or temperature variation) conditions. Our results are not especially sensitive
to whether the bracket terms are included in the limiting process, generally agreeing
with the simpler procedure ($F = f_E$), particularly at higher resolutions, so long as the
constant coefficient $f_E$ is adjusted accordingly.
}

Apart from the Newtonian (versus general relativistic) MHD
treatment, the numerical methods concerning nucleosynthesis, reactant solvers, energy coupling, etc., are identical
to our previous TDE studies, so we refer the reader to the publications listed above for details not provided here.

\subsection{Magnetic Fields}
\label{subsec:magnetic}

A few of our models include either toroidal or poloidal magnetic fields scaled
globally to a specified strength parameter $\beta$ such that the ratio of magnetic
to thermal pressures everywhere satisfies $P_B/P \le \beta$.
\added{This scaling is applied only when initially mapping the stars to the grid, allowing
the field amplitudes to amplify over time.}
The magnetic field inside the star (defined by a threshold pressure ratio $\delta_P \equiv P/P_{max} = 10^{-5}$ of the central peak $P_{max}$) 
is calculated from a single-component vector potential
\begin{alignat}{2}
A^i &= r_{cyl}^{\alpha_1} e^{-((r-r_c)/(\alpha_2 R_*))^2} && \quad \text{if $\delta_P > 10^{-5}$}  \nonumber \\
    &= 0 && \quad \text{otherwise} ~,
\label{eqn:magVP}
\end{alignat}
where $r$ is a spherical coordinate, $r_{cyl}$ the cylindrical coordinate, $r_c$ the star centers, $R_*$ the star radii,
and the parameters $\alpha_i$ are tuned to generate a desired field strength at the
stellar surface once a strength parameter is chosen to set the maximum field amplitude in the core.
We take $(\alpha_1,~\alpha_2) = (0,~0.28)$ for toroidal fields, and (4,~0.25) for poloidal.
The non-zero poloidal field variable $\alpha_1$ prevents
excessively steep gradients of the magnetic pressure along the pole, and effectively describes a current loop
potential with a radius equal to roughly half the star radius.
The index on $A^i$ represents an azimuthal component ($A^i \equiv A^\phi$) for poloidal fields,
and a longitudinal one ($A^i \equiv A^z$)for toroidal.

\subsection{1D Reactive Shocks}
\label{subsec:params_1d}

The one-dimensional test problems proposed by 
\citet{Katz19} are a set of planar shock tube configurations with the following initial conditions:
uniform $5\times10^6$ g/cm$^3$ density, $10^7$ K temperature, $-2\times10^8$ cm/s velocity,
and $-1.1\times10^8$ cm/s$^2$ acceleration, across a grid of length $1.64\times10^9$ cm.
The left boundary (at $x=0$) is a rigid reflecting wall. The right boundary is open with
symmetric conditions allowing gas to flow perpetually onto the grid subject to gravitational acceleration.
The compositions are initialized to either equal mass (50/50) carbon/oxygen,
or 5/50/45 mass fractions He/C/O.
Unlike the 2D models discussed in the next section, we do not use geometric zoning for these tests.
Instead the entire grid is covered uniformly with evenly spaced cells since ignition is
expected to trigger somewhere off the reflective boundary as accreting material
powers the outward propagating shock front. 
In each case ignition is defined at the time and location where the gas temperature first reaches $T=4\times10^9$ Kelvin.

\subsection{2D WD Collisions}
\label{subsec:params_2d}

\begin{figure}
\hspace{1.2in}\includegraphics[width=0.6\textwidth]{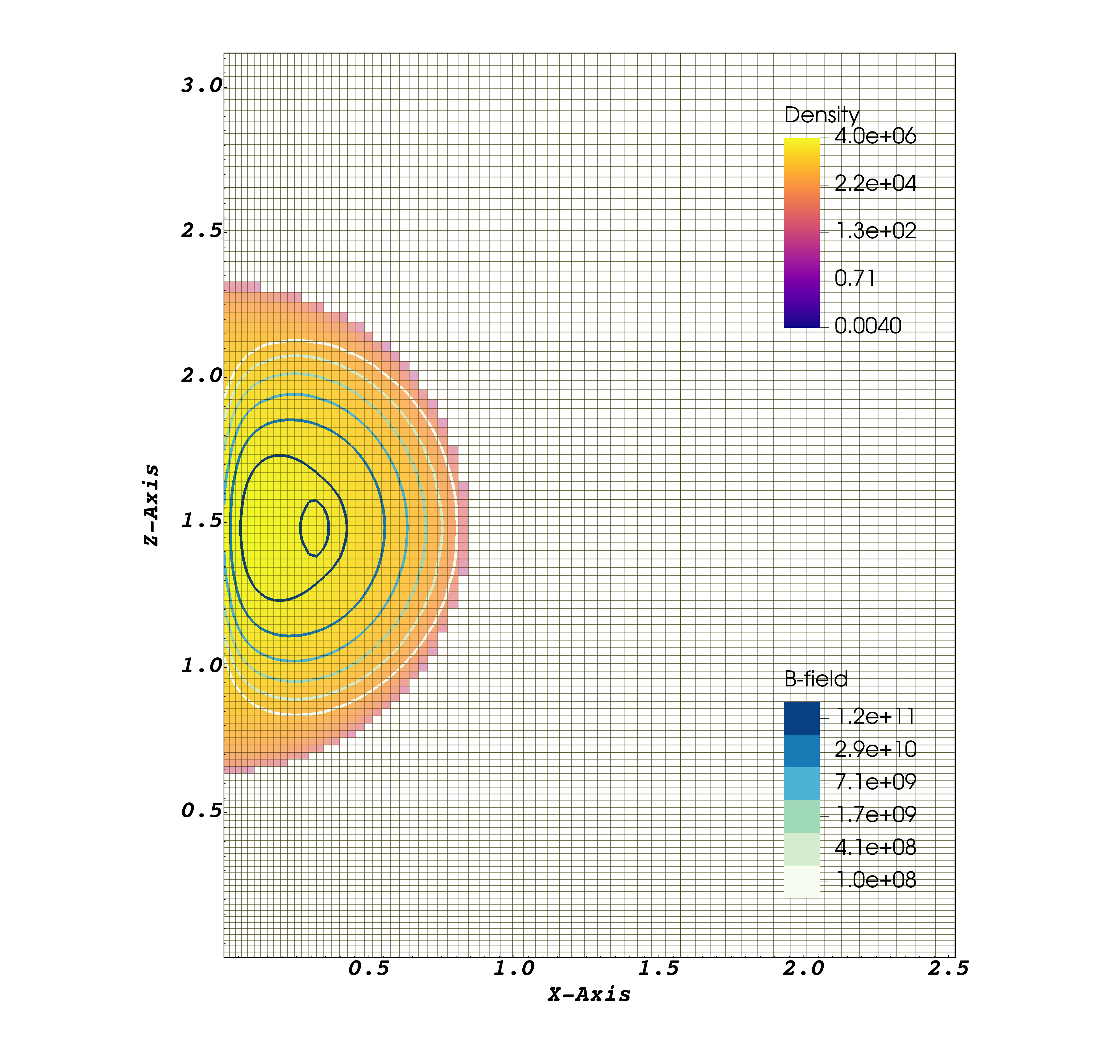}
\caption{
Sample grid demonstrating cylindrically symmetric coordinate lines geometrically converging
onto both the equatorial plane 
\added{
where reflection boundary conditions are imposed
}
($z=0$) and pole axis ($x=0$). All collision calculations are run on similar
grids but at much higher resolutions set by the maximum cell resolution and zone count along each axis.
For clarification purposes, this particular image was generated at 200 km resolution.
Also shown are the gas density in cgs units (with a color map) and the initial magnetic field 
amplitude contours in units of Gauss for the poloidal configuration.
Axis labels are in units of $10^4$ km.
}
\label{fig:geomgrid}
\end{figure}

\replaced {
Our TDE convergence studies \citep{Anninos22} managed to cover more than five decades in spatial scales,
achieving sub-kilometer resolutions at initiation sites near the equatorial plane
by using static geometric zoning (not adaptive refinement).
This strategy increases cell dimensions along an axis orthogonal to the burn front
by the product $\Delta z_{i+1} = (1 + \epsilon) \Delta z_i$, where the geometric constant ($\epsilon > 0$)
depends on the number of zones covering the length of the axis, and is recovered by bisection from
} {
To achieve a dynamical range broad enough to cover the nearly five decades in spatial scales
needed to address stability and convergence uncertainties, we apply a static geometric
zoning strategy similar to \cite{Anninos22}, but tuned to concentrate mesh cells near ignition sites
and aligned (approximately) with developing shock structures near the polar and equatorial planes.
This strategy increases cell dimensions along both axes
by the product $\Delta z_{i+1} = (1 + \epsilon) \Delta z_i$, where the geometric constant ($\epsilon > 0$)
depends on the number of zones covering the length of the axis (demonstrated here for
the $z$-axis), and is recovered by bisection from
}
\begin{equation}
\Delta z_{\text{min}} = \frac{\epsilon ~L_z}{(1+\epsilon)^{N_z} - 1} ~,
\end{equation}
given a target resolution $\Delta z_{\text{min}}$ (smallest cell width), number of zones $N_z$, and grid half-length $L_z$.
\replaced{
A similar geometric refinement procedure is easily adapted to the nearly planar shock configuration
expected from the head-on collision of two WD stars. Grids are constructed 
as shown in Figure \ref{fig:geomgrid} with mesh lines
converging onto the centroid of the binary system along the trajectory-aligned axis (the $z$-axis), as well
as the transverse direction along the equatorial plane (the $x$-axis projected from cylindrical coordinates).
} {
A typical grid constructed in this fashion is shown in Figure \ref{fig:geomgrid} with mesh lines
converging onto the centroid of the binary system along the trajectory-aligned axis (the $z$-axis), as well
as the transverse direction along the equatorial plane (the $x$-axis projected from cylindrical coordinates).
}
We maintain separate and distinct geometric factors to achieve desired (but independent) resolutions along each axis,
with typical values $\epsilon \approx 10^{-2}$ limiting cell-to-cell size progressions to roughly $1$\%.

Initially the horizontal extent of the grid (along the transverse direction) is fixed at three times the maximum
radius of the two stars. The vertical grid half-length $L_z$ (along the collision trajectory) is also
fixed (initially), but allowing for unequal masses, is adjusted to the separation distance between
centers of mass $d_0$ as
\begin{equation}
L_z = \text{max} \left( 3 R_{*} + d_0 \frac{M_T - M_*}{M_T} \right) ~,
\end{equation}
where $R_{*}$ are the individual star radii, $M_*$ are the stellar masses, $M_T$ is the total binary mass,
\added{
and the max() operation is performed over both stars in case of unequal masses.
}
The stars are positioned either at near contact (with centers separated by a factor of 1.1 times the sum of the radii,
or just beyond the L1 Lagrange point (separated by several star radii), with assigned freefall velocities
\begin{equation}
v_0 = \pm \frac{M_T - M_*}{M_T} v_{ff} = \pm \frac{M_T - M_*}{M_T} ~\sqrt{\frac{2GM_T}{d_0}} ~.
\end{equation}

A summary of grid and model parameters is provided in Table \ref{tab:runs_2d}, including masses, compositions,
grid resolutions, initial separations ($d_0$), initial velocities ($v_0$), and magnetic field
amplitudes in the core ($|B|_{max}$) and averaged along the surface ($|B|_{surf}$). The number of
zones covering the grid varies from $N_x \times N_z = 112 \times 224$ at the coarsest (100 km) resolution to
$448\times 896$ at the finest (1 km). 
\added{
The greater number of zones along the $z$-axis allows us to maintain
roughly equivalent cell resolutions when the vertical extent of the grid is increased to accommodate
large initial separations. This also effectively provides greater cell resolution some distance off the equatorial
plane by reducing the geometric (cell-to-cell refinement) ratio.
}

In addition to geometric zoning, we allow the mesh to move and follow debris ejected after impact.
Mesh velocity (or ALE remap) is triggered only
when stellar matter is detected near the grid boundaries, and then set to
\begin{equation}
\dot{x}^i_g = \zeta^i ~\overline{v}^i ~\frac{x^i - x^i_c}{x^i_{o} - x^i_c} ~,
\end{equation}
where $\dot{x}^i_g$ is the grid velocity assigned to each cell node in vector index notation, $\zeta^i$ is a constant different along each axis
but adjusted in time based on the proximity of material to grid boundaries (with typical values between 1 and 2),
$\overline{v}^i$ is the mass weighted fluid velocity,
$x^i$ is the grid coordinate, $x^i_c$ is the mass centroid, and $x^i_{o}$ is the outer grid boundary.
This formulation responds to the mass-weighted Lagrangian fluid velocity while respecting geometric spacing
and at the same time preventing nodes from punching through
their neighbors by linearly distributing the mesh velocity outward from the centroid containing the highest mesh concentration.

Symmetric boundary conditions are imposed at all open boundaries, allowing the possibility of significant outflow at late times.
In addition, a high order multipole expansion for the gravity potential adjusts to evolving tidal forces,
and reflection boundary conditions across both polar and equatorial (for equal mass collisions) saves computational time.
\added{
The multipole boundary conditions are centered on the mass centroid and include moments up to 10th order.
}
Outside the stars we impose a background density of $10^{-8}\rho_{\text{max}}$ and pressure $10^{-5} p_{\text{min}}$,
where $\rho_{\text{max}}$ and $p_{\text{min}}$ are the maximum core density and minimum pressure at the surface
of both stars. 
\added{
Though they play no role in the evolutions, we maintain these floor values throughout the calculations.
The calculations are terminated when the relative nuclear energy production
(defined as the ratio of energy generated per cycle to the total accumulated energy produced) drops below 0.5\%.
}

\begin{deluxetable}{lcccccccccc}
\tablecaption{Collision Parameters \label{tab:runs_2d}}
\tablewidth{0pt}
\tablehead{
\colhead{series}          & 
\colhead{mass}            & 
\colhead{composition}     & 
\colhead{he4}        & 
\colhead{c12}        & 
\colhead{o16}        & 
\colhead{resolution}      & 
\colhead{$d_0$}           & 
\colhead{$v_0$}           & 
\colhead{$|B|_{surf}$}    &
\colhead{$|B|_{max}$}     \\
                          & 
($M_\odot$)               & 
                          & 
($M_\odot$)               & 
($M_\odot$)               & 
($M_\odot$)               & 
(km)                      & 
($10^4$ km)               & 
(km/s)                    & 
(Gauss)                   &
(Gauss)             
}
\startdata
M1a     & 1.0  & C/O    & -     & 0.52 & 0.52 & 1-100  &  1.8  & 2100   & -    & -    \\
M1b     & 1.0  & Mesa   & 0.001 & 0.43 & 0.60 & 1-100  &  1.8  & 2100   & -    & -    \\
M6a     & 0.6  & C/O    & -     & 0.31 & 0.31 & 1-100  &  1.8  & 2100   & -    & -    \\
M6b     & 0.6  & C/O    & -     & 0.31 & 0.31 & 1-100  &  3.2  & 1600   & -    & -    \\
M6c     & 0.6  & C/O    & -     & 0.31 & 0.31 & 1-100  &  3.2  & 1600   & $6\times10^7$ & $1\times10^{11}$  \\
M6d     & 0.6  & C/O    & -     & 0.31 & 0.31 & 1-100  &  1.7  & 2200   & $1\times10^8$ & $5\times10^{10}$  \\
M6e     & 0.6  & Mesa   & 0.01  & 0.26 & 0.34 & 1-100  &  1.7  & 2200   & -    & -    \\
M6f     & 0.6  & C/O    & -     & 0.24 & 0.37 & 1-100  &  1.7  & 2200   & -    & -    \\
M6g     & 0.6  & He/C/O & 0.05  & 0.28 & 0.28 & 1-100  &  1.7  & 2200   & -    & -    \\
\enddata
\end{deluxetable}

\section{Results}
\label{sec:results}

\subsection{1D Reactive Shocks}
\label{subsec:burntube}

\begin{figure}
\hspace{1.5in}\includegraphics[width=0.5\textwidth]{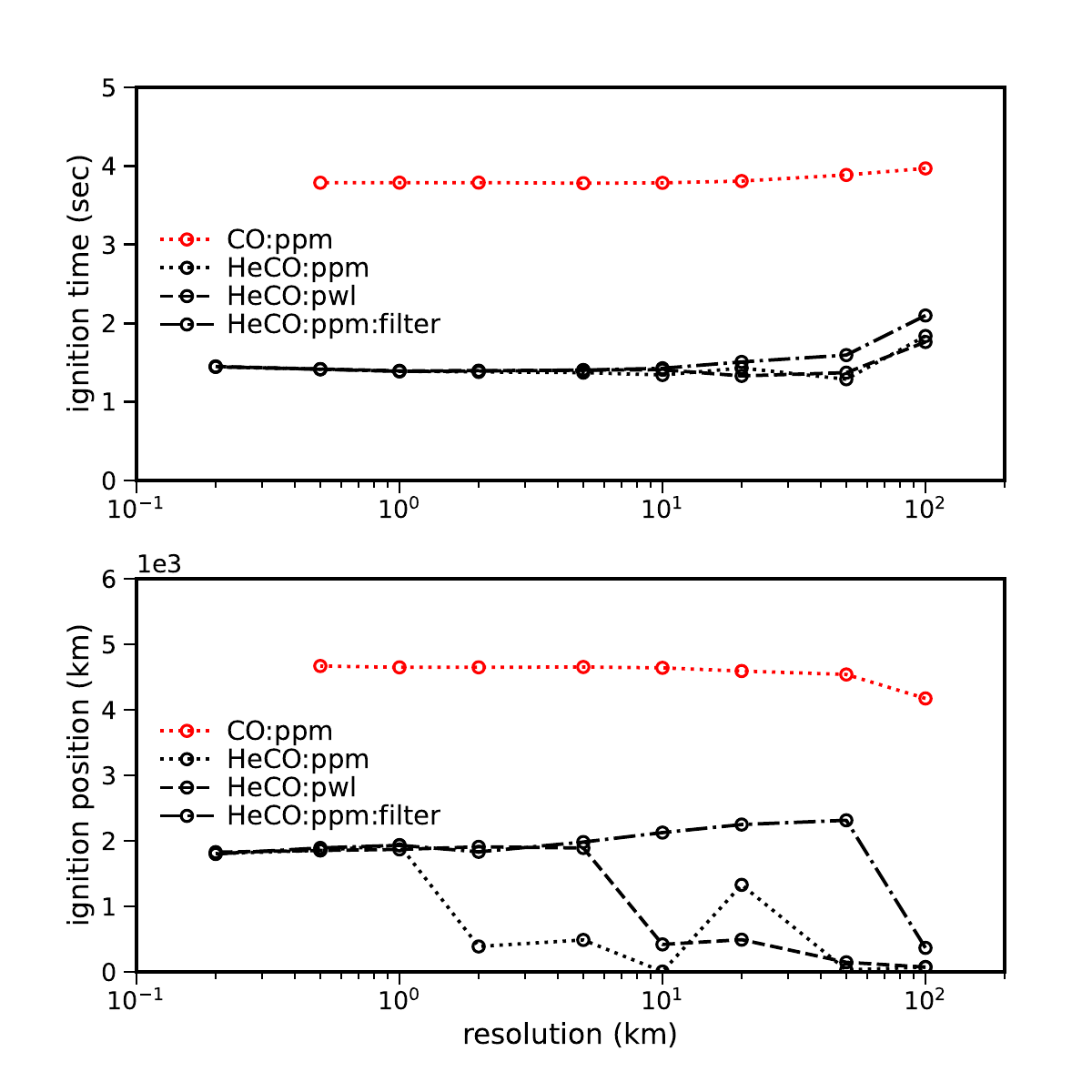}
\caption{
Time of ignition (top) and location of initiation point (bottom) from the 1D reactive shock tests
plotted as a function of grid resolution with a fixed energy limiter $f_E=0.04$. Red curves correspond to
the pure C/O test, black curves to the He/C/O test with each line type
representing results from the different solver options discussed in the text.
Ignition is defined at the time and place where the temperature first exceeds $4\times10^9$ K.
}
\label{fig:btubedx_elim}
\end{figure}

\begin{figure}
\hspace{1.5in}\includegraphics[width=0.5\textwidth]{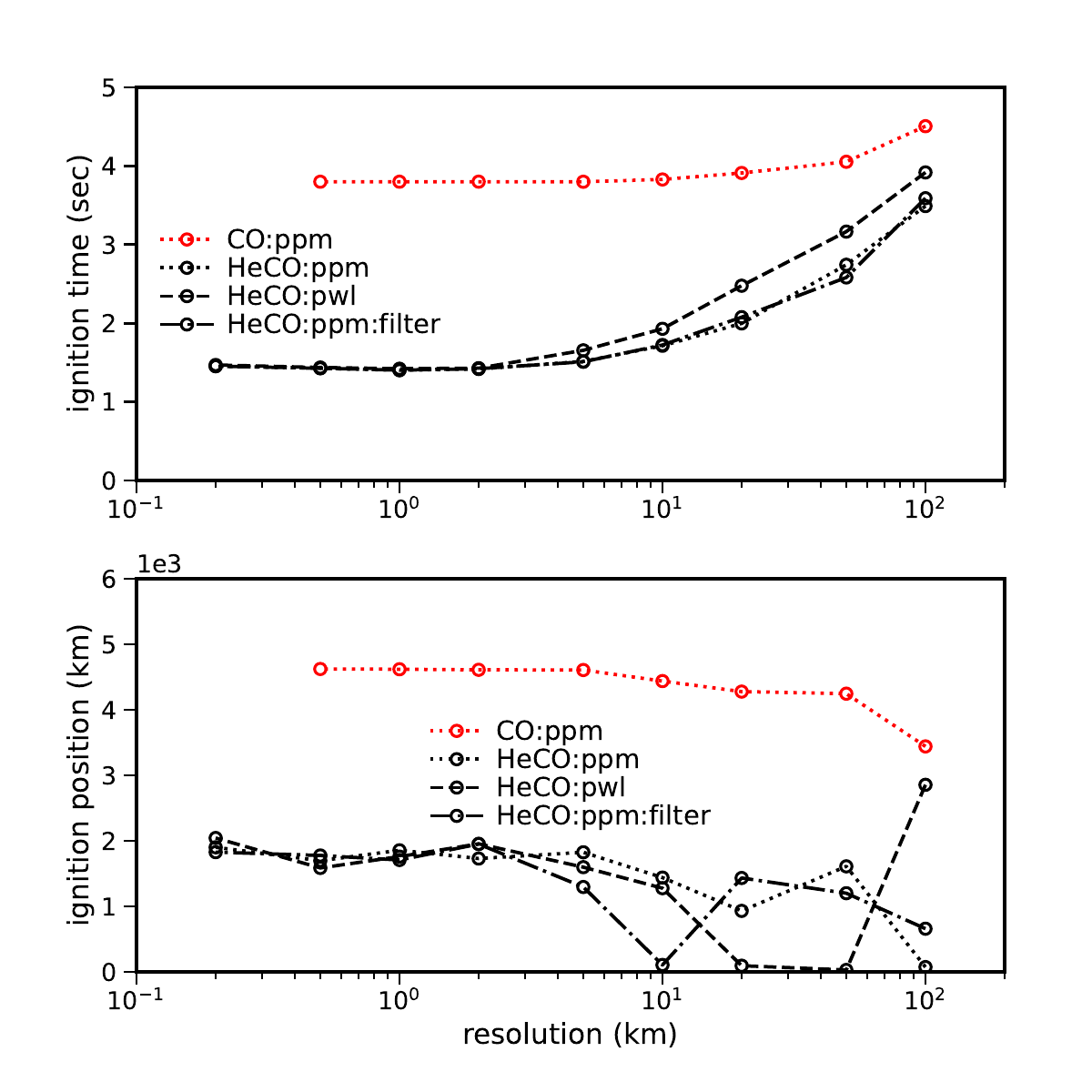}
\caption{
As Figure \ref{fig:btubedx_elim} except for a temperature limiter $f_T=0.01$.
}
\label{fig:btubedx_tlim}
\end{figure}

Figures \ref{fig:btubedx_elim} and \ref{fig:btubedx_tlim} compare ignition times and locations as a function of
cell resolution from the two different composition
tests (pure C/O shown in red, He/C/O in black), run with three different hydrodynamics solver
options: second order piecewise linear flux reconstruction (labeled `pwl'), third order piecewise parabolic
reconstruction (labeled `ppm'), and filtered piecewise parabolic (`ppm:filter').
The third `ppm:filter' option uses third order flux reconstruction,
but, in order to compare our results against KK20, 
additionally applies a grid-based filter to suppress nuclear reactions
in the first zone off the reflection boundary.
These options are applied across nine grid resolutions ranging from 100 to 0.2 km,
and repeated for two versions of the burn limiter: one limits the energy generation and deposition
to 4\% of the internal fluid energy ($f_E \equiv |\delta e_{nuc}|/e = 0.04$), and the second constrains temperatures
to 1\% relative change cycle to cycle ($f_T \equiv |\delta T|/T = 0.01$). 
In both versions, limiting is enforced with an iterative bisection
procedure that converges on a local timestep satisfying the imposed constraints.

Several important conclusions can be drawn from these results. First, although we have not presented
calculations of the C/O test with different hydro options, we note that the displayed ppm results
are identical to both the pwl and ppm+filter options.
The C/O test additionally appears largely insensitive to resolutions below 20 km, where
the ignition time and location curves flatten at $\sim 3.8$ sec and $\sim 4600$ km to better than a few percent,
\added{
in good agreement with Figure 1 of KK20.
}

Next, our results for the He/C/O case are generally consistent with both
KZ19 and KK20 in that we observe different behaviors depending on grid resolution.
At coarse resolutions ($\gtrsim 10$ km) initiation apparently triggers at random locations.
At $5$ km or better (depending on hydro solver options and limiter procedures)
the initiation site converges to about 1900 km from the boundary at $\sim1.4$ sec,
\added{
in reasonable agreement with KK20, but differing from KZ19 by about a factor of two.
It is not clear why our results (and those of KK20) differ from KZ19 by such an amount, other than
perhaps due to the simplified 13-isotope network used by KZ19, or to
different numerical solvers.
}
KK20 attribute the unstable behavior at coarse resolution to fast reaction rates fueling artificial hotspots
seeded by numerically generated sound waves at the reflection boundary. They further demonstrate
this numerical instability can be suppressed with a burn filter (mask) to prevent nuclear reactions
at the reflection boundary, or by decreasing the artificially generated hotspot width by increasing resolution.
Figures \ref{fig:btubedx_elim} and \ref{fig:btubedx_tlim} confirm both conjectures. 
Comparing the dotted curve against the dot-dashed in Figure \ref{fig:btubedx_elim}, we
observe the filter increases stability by nearly two orders of magnitude as measured by the resolution at which the convergent plateau
first develops (at 1, 5, and 50 km for the ppm, pwl, and filtered calculations respectively).
Interestingly we do not observe the same benefit by masking nuclear reactions
when limiting the temperature in place of energy. In that case, as shown in Figure \ref{fig:btubedx_tlim},
all three hydro options behave similarly at less than 10 km, and converge nicely 
somewhere between 2 to 5 km.

\begin{figure}
\hspace{1.5in}\includegraphics[width=0.5\textwidth]{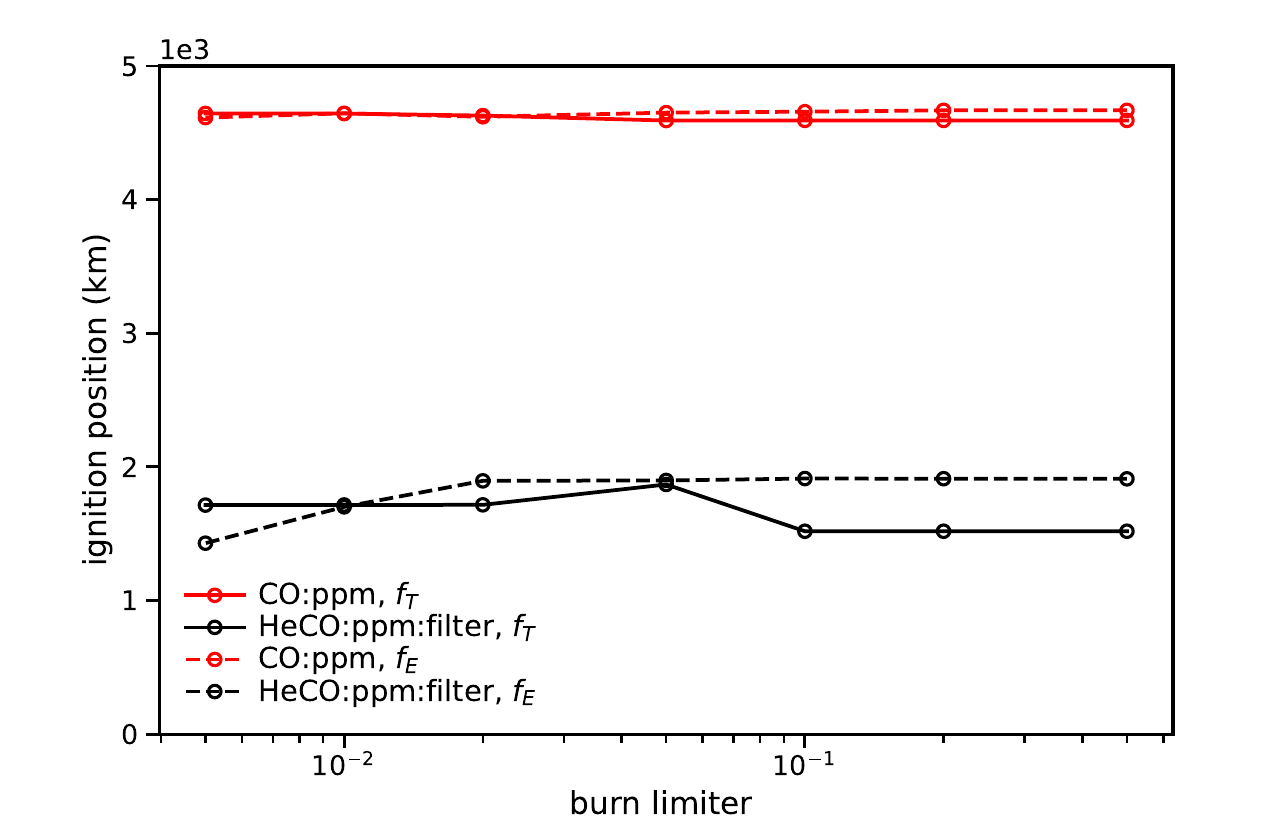}
\caption{
Ignition position is plotted as a function of the burn limiter amplitude ($f_T$ or $f_E$) at fixed 1 km grid resolution.
Solid (dotted) lines represent results by limiting temperature (energy).
}
\label{fig:btubelim}
\end{figure}

Considering the importance of burn limiters, we present in Figure \ref{fig:btubelim}
parameter studies demonstrating the effects of limiting either temperature or energy.
The calculations were carried out at fixed 1 km resolution where results 
(from Figures \ref{fig:btubedx_elim} and \ref{fig:btubedx_tlim}) appear well-resolved, while
varying both limiter parameters ($f_T, ~f_E$) from 0.005 to 0.5.
Apart from a slight break in the He/C/O test at $f_T=0.05$, Figure \ref{fig:btubelim} shows little
sensitivity to $f_T$ (roughly 1\% and 10\% variance in the C/O and He/C/O tests respectively) 
across two orders of magnitude.
On the other hand, the energy limiter (in Figure \ref{fig:btubelim}) gives consistent results at $f_E > 0.01$,
but begins to adversely affect the solution at smaller values as nuclear energy is increasingly neglected each cycle.
Overall it is encouraging to see the different hydro methods and limiting procedures converge at roughly 5 km,
so long as we restrict limiter amplitudes to $0.01 \le (f_T, ~f_E) \le 0.05$.

\subsection{Axisymmetric WD Collisions}
\label{subsec:wdcollision}

\subsubsection{Unmagnetized}
\label{subsubsec:unmagnetized}

\begin{figure}
\hspace{1.5in}\includegraphics[width=0.5\textwidth]{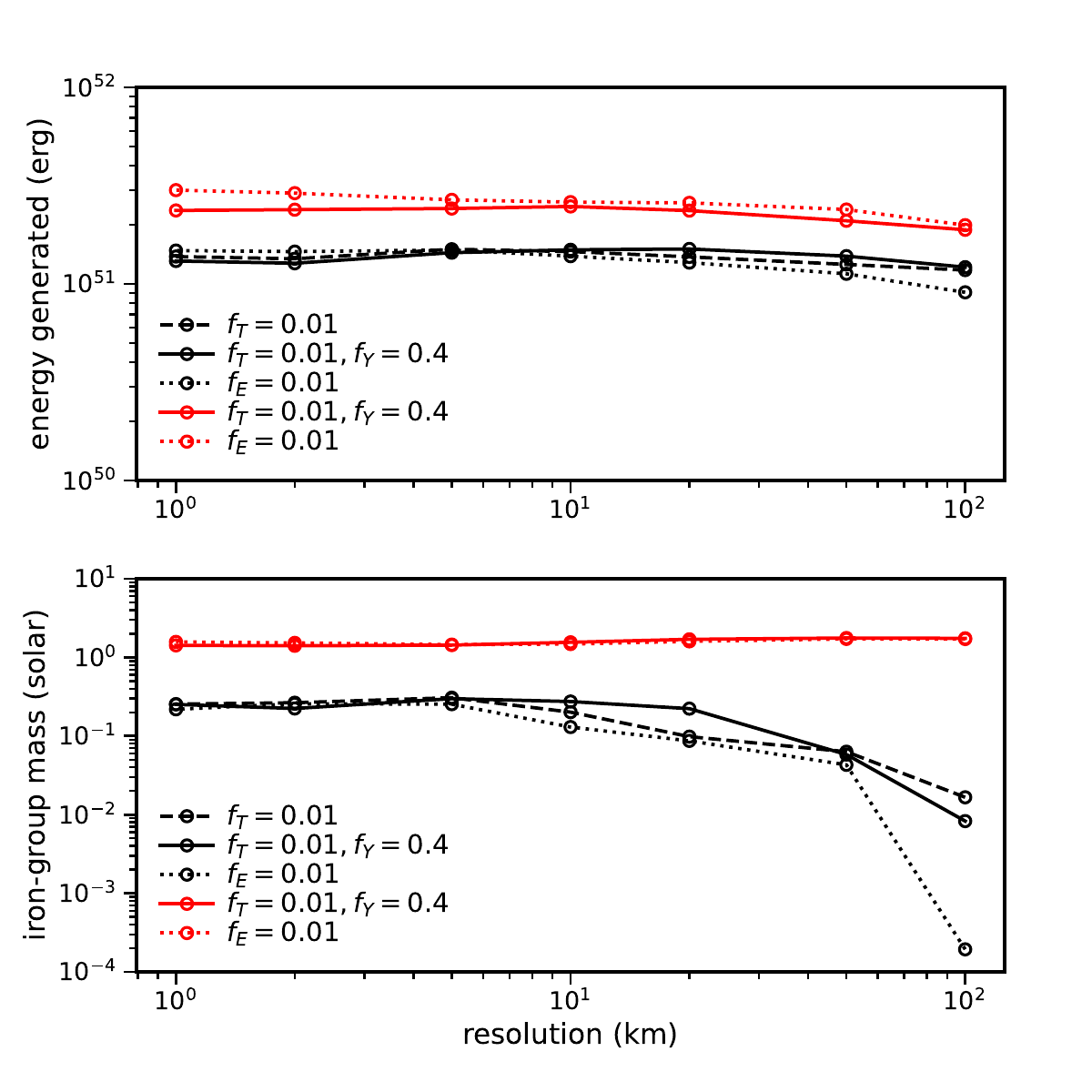}
\caption{
Total nuclear energy generated (top) and final iron-group mass (bottom) as a function of grid resolution.
Black (red) represents results from the collision of two 0.6 (1.0) $M_\odot$ stars.
The various line types correspond to different implementations of limiters as indicated in the figure legend
and discussed in the text. The iron-group mass sums densities from the
production of $^{44}$Ti, $^{48}$Cr, $^{52}$Fe, $^{54}$Fe, and $^{56}$Ni.
}
\label{fig:wdwdres}
\end{figure}

Figure \ref{fig:wdwdres} plots the total released nuclear energy and iron-group mass 
as a function of grid resolution resulting from the collision of two 0.6 $M_\odot$ (in black)
and two 1 $M_\odot$ (in red) WDs with initially uniform 50/50 C/O compositions (models M1a and M6a).
Here and throughout, iron-group is defined as the sum of the heaviest elements in our network:  
$^{44}$Ti, $^{48}$Cr, $^{52}$Fe, $^{54}$Fe, and $^{56}$Ni.
Seven sets of calculations were performed for each interaction at resolutions between 100  to 1 km,
and for three different burn limiter options: (1) constraining cyclic temperature variations with $f_T=0.01$,
(2) constraining energy generation and deposition with $f_E=0.01$,
and (3) limiting both temperature and reactant relative changes with $f_T=0.01$ and $f_Y\equiv \delta Y/Y = 0.4$ respectively.
Notice the nuclear energy generated is relatively insensitive to grid resolution or limiter strategies, varying by less than 40\% 
over the entire parameter space and less than 7\% at cell sizes $\le 10$ km.
Iron-group mass, on the other hand, is significatly more sensitive to grid resolution, limiter options,
and collision parameters (represented by the two WD masses). Although the heavier mass impact
is fairly robust (displaying only a modest average downward trend of about 0.003 $M_{\odot}$/km),
results from the 0.6 $M_\odot$ case
vary roughly by factors of two between 10 and 50 km cell sizes, but then appear to stabilize and converge
to within 10\% below 5 km.

\begin{figure}
\hspace{1.5in}\includegraphics[width=0.5\textwidth]{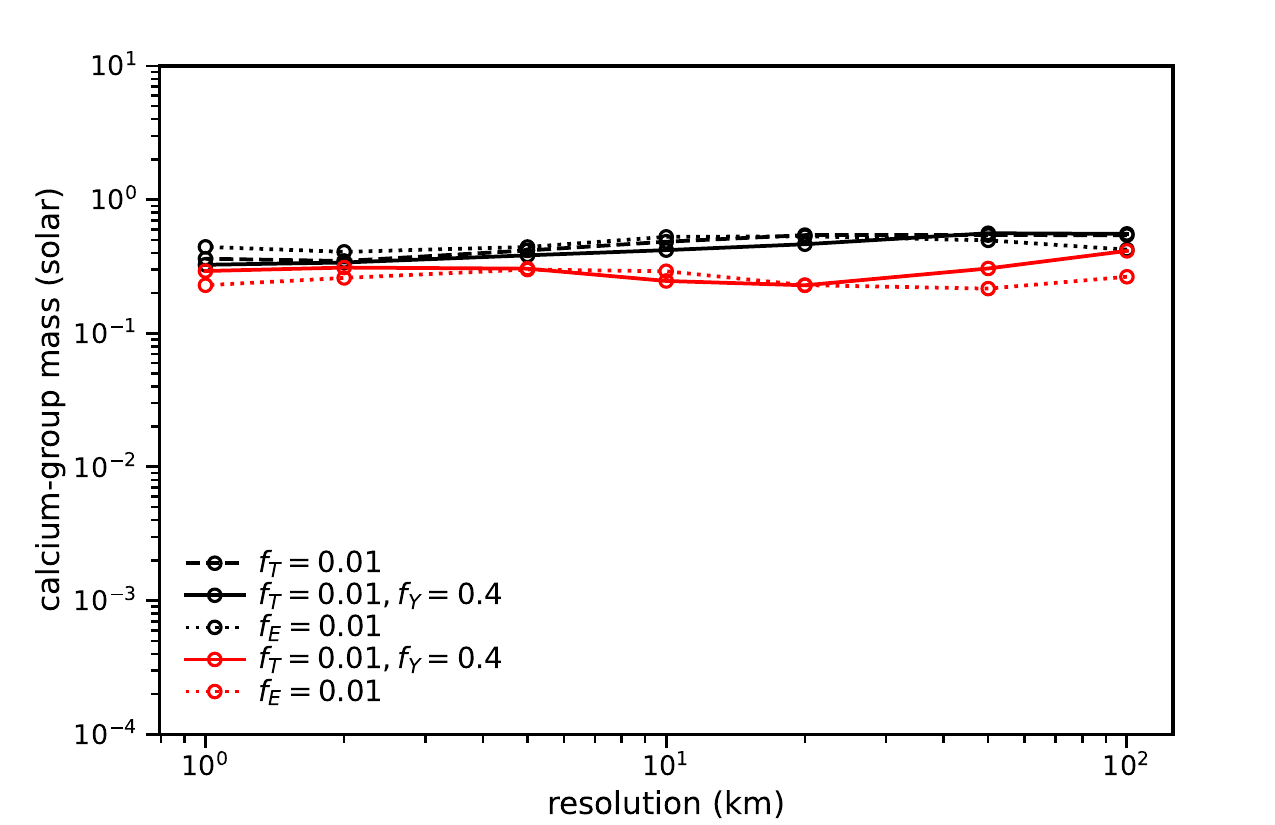}
\caption{
Total calcium-group mass (including $^{28}$Si, $^{32}$S, $^{36}$Ar,and $^{40}$Ca) is plotted as a function of grid resolution.
Black (red) curves represent results from the collision of two 0.6 (1.0) $M_\odot$ stars.
}
\label{fig:wdwdres_ca}
\end{figure}

We find similar converging trends for intermediate mass elements (IMEs) in Figure \ref{fig:wdwdres_ca}, where we plot
the total calcium-group mass in the bottom plate
(including $^{28}$Si, $^{32}$S, $^{36}$Ar, and $^{40}$Ca). 
\added{
Although we find these elements exhibit somewhat greater sensitivity to limiter
parameters (20\% at 5 km, roughly double the iron-group uncertainties), they appear to be
less sensitive to grid resolution, remaining robust (basically plateaued) even at resolutions
as coarse as 50 km.  There is, however, a slight trend towards greater intermediate mass production
at coarse resolutions. This effect, together with the fact that total energy released is dominated
mostly by silicon production, following the same growth pattern and accounting for more than
half the total energy before nickel forms in quantity, accounts for the insensitivity of
total energy released (in Figure \ref{fig:wdwdres}) to nickel production.
Even at our highest resolution, the mass fraction of silicon exceeds that of nickel,
and any loss of energy due to the lack of iron-group production at coarse resolutions 
is essentially made up by making more IMEs.
}

\begin{figure}
\hspace{1.5in}\includegraphics[width=0.5\textwidth]{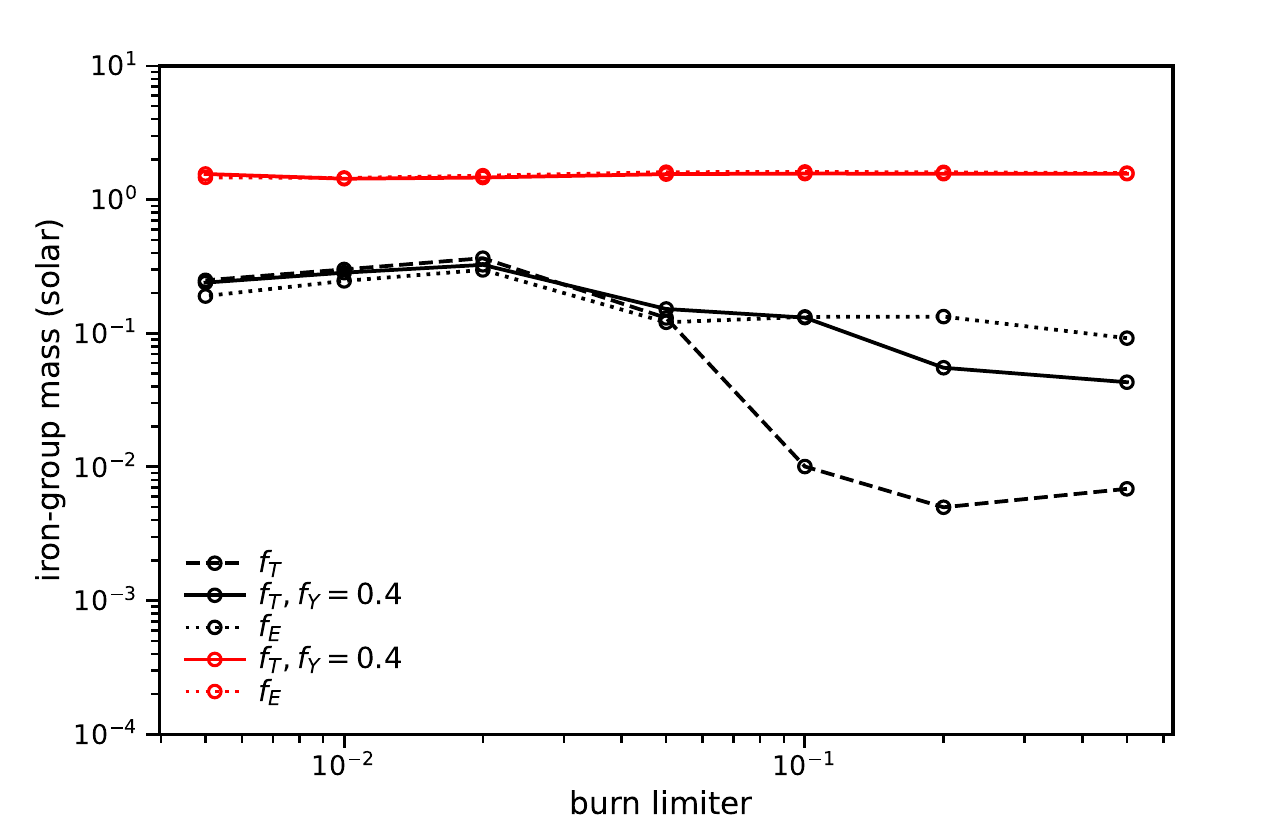}
\caption{
Total iron-group mass from the the collision of two 0.6 $M_\odot$ (black) and 1 $M_\odot$ (red) WDs
is plotted as a function of burn limiter amplitude ($f_T$ or $f_E$).
All calculations were run at 5 km resolution.
}
\label{fig:wdwdlim}
\end{figure}

Burn limiter sensitivities are demonstrated in Figure \ref{fig:wdwdlim} for the iron-group mass plotted
as a function of limiter amplitude ($f_T$ or $f_E$).
All calculations in the figure are performed at the same grid resolution of 5 km,
determined by Figure \ref{fig:wdwdres} to be reasonably well converged.
Although we observe the energy generated to be largely insensitive to limiter procedures
or parameters, the iron-group mass on the other hand exhibits order of magnitude uncertainties
when allowed fractional variations (in either temperature or energy) are relaxed to 10\% or greater.
We additionally point out that the energy limited iron-group mass (dotted line in the bottom plate) drops at a slightly
faster rate than either of the temperature limiter curves (solid and dashed lines),
suggesting values $f_E < 0.01$ are too restrictive and begin
to impact the hydrodynamics in a manner similar to what is observed in Figure \ref{fig:btubelim}
from the 1D shocktube test.

On the whole, these results suggest that over the parameter space considered here ($0.005 \le (f_T, ~f_E) \le 0.5$), 
there is a lower bound on $f_E$ below which energy
release is visibly affected, and an upper bound on $f_T$ where the two limiting procedures begin to stray from
each other resulting in unreliable behavior.
Values around 0.01-0.03 represent a reasonable compromise between reliability and
consistency for both energy and temperature limiters, and result in well behaved monotonically converging
solutions with just 10\% uncertainty at 5 km resolution.
Table \ref{tab:results_2d} summarizes our results by tabulating the total nuclear energy released
($e_{nuc}$), iron-group mass ($M_{Fe}$), and calcium-group mass ($M_{Ca}$) using different limiter
parameters at 1 km and 5 km resolutions.

\begin{deluxetable}{lccccccc}
\tablecaption{Collision Results  \label{tab:results_2d}}
\tablewidth{0pt}
\tablehead{
\colhead{Series}          & 
\colhead{$f_E$}           & 
\colhead{$f_T$}           & 
\colhead{$f_Y$}           & 
\colhead{Resolution}      & 
\colhead{$e_{nuc}$}       & 
\colhead{$M_{Fe}$}        &
\colhead{$M_{Ca}$}        \\
                          & 
                          & 
                          & 
                          & 
(km)                      & 
($10^{51}$ ergs)          & 
($M_\odot$)               &
($M_\odot$)              
}
\startdata
M1a      &  0.01 & -    & -    & 1(5)  &  $3.0(2.7)$  & 1.57(1.45) & 0.42(0.55)      \\
''       &  -    & 0.01 & 0.4  & 1(5)  &  $2.4(2.4)$  & 1.42(1.43) & 0.54(0.55)      \\
M1b      &  -    & 0.04 & -    & 1(5)  &  $1.8(1.7)$  & 1.42(1.52) & 0.30(0.27)      \\
M6a      &  0.01 & -    & -    & 1(5)  &  $1.5(1.5)$  & 0.22(0.25) & 0.75(0.78)      \\
''       &  -    & 0.01 & -    & 1(5)  &  $1.4(1.5)$  & 0.25(0.31) & 0.59(0.71)      \\
''       &  -    & 0.01 & 0.4  & 1(5)  &  $1.3(1.4)$  & 0.25(0.30) & 0.56(0.69)      \\
M6b      &  0.01 & -    & -    & 1(5)  &  $1.5(1.4)$  & 0.24(0.20) & 0.40(0.48)      \\
M6c      &  0.01 & -    & -    & 1(5)  &  $1.4(1.4)$  & 0.23(0.20) & 0.39(0.47)      \\
M6d      &  0.01 & -    & -    & 1(5)  &  $1.4(1.4)$  & 0.29(0.23) & 0.33(0.42)      \\
M6e      &  -    & 0.04 & -    & 1(5)  &  $0.9(0.9)$  & 0.03(0.03) & 0.44(0.57)      \\
M6f      &  -    & 0.04 & -    & 1(5)  &  $0.9(0.8)$  & 0.04(0.02) & 0.55(0.54)      \\
M6g      &  -    & 0.04 & -    & 1(5)  &  $0.8(0.9)$  & 0.04(0.05) & 0.42(0.56)      \\
\enddata
\end{deluxetable}

\subsubsection{Magnetized}
\label{subsubsec:magnetized}

In this section we study the convergence of nuclear reactions from colliding 0.6 $M_\odot$ magnetized white dwarf stars.
We consider both toroidal (series M6c) and poloidal (series M6d) configurations with
parameters tabulated in Table \ref{tab:runs_2d}. Field amplitudes are normalized by
enforcing $\beta=0.03$ in the core and choosing $\alpha_i$ in equations (\ref{eqn:magVP})
which produce fields in the central core that are within theoretical bounds, and surface amplitudes that sample the high end of observations 
\citep{Franzon15,Putney95,Schmidt95}. For this purpose we define surface amplitude to mean the field strength at a radial distance of $0.9~R_*$.

\begin{figure}
\hspace{-0.2in}\includegraphics[width=1.5\textwidth]{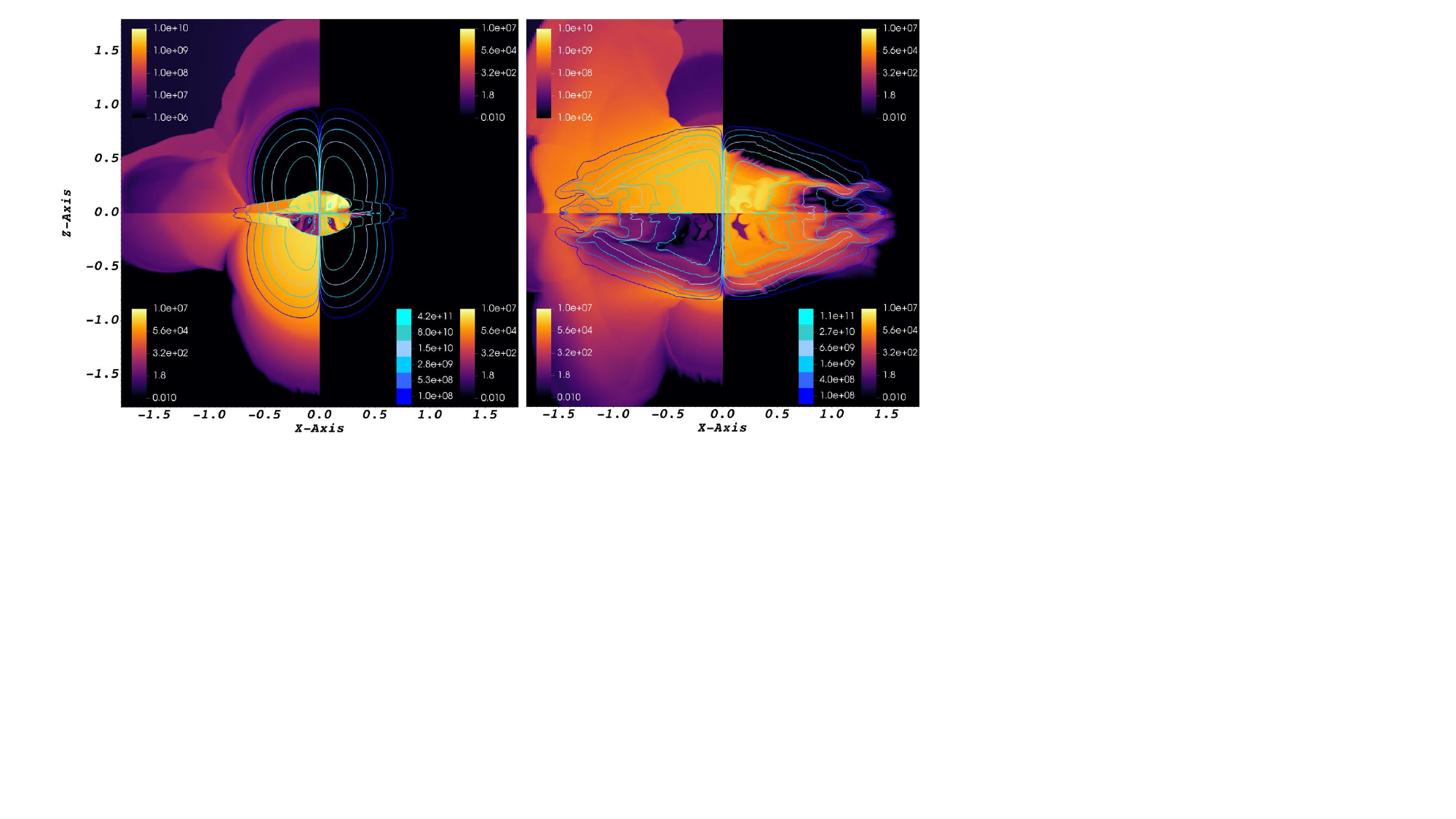}
\vspace{-2.7in}\caption{
Evolution sequence from the magnetized collision of two 0.6 $M_\odot$ stars, showing 
in each of the plates the temperature (top left),
carbon density (bottom left), nickel density (top right), and calcium density (bottom right).
\added{
Our calculations impose reflection (and polar) boundary conditions along the
equatorial plane (and pole axis) - we do not simulate the entire domain, as perhaps these plates
might suggest, but merely stack images of the different fields into different quadrants for
visualization purposes.
}
Also shown are line contours of the (toroidal) magnetic field amplitude in Gauss units.
The left plate corresponds to an early time (6.6 s) immediately following the impact, shock formation, and detonation
as the burn front begins to move radially outward from the initiation site; the right plate to a
later time (7.4 s) when the burn front reaches the opposite surface along the pole
and heavy elements eject along the collision plane (right).
Axes labels are in units of $10^4$ km.
}
\label{fig:plate_Btor}
\end{figure}

Figure \ref{fig:plate_Btor} shows two images from the toroidal field case (M6c) at 2 km resolution
taken at two different times: 6.6 secs from the start of the simulation (left), and 7.4 (right).
The left plate corresponds to a time immediately after impact when the collisional shock is well established
at about $2\times10^3$ km in the vertical direction as it begins to propagate through the star.
Nearly a second later in the right image the burn front has traversed the entire star, reaching the
opposite surface along the pole as the heavy elements formed in its wake are ejected along the $x$-axis.
Each plate is composed of four quadrants imaging the temperature (top left), nickel density (top right), 
carbon density (bottom left, providing a measure of unburnt gas), 
and calcium density (bottom right), all in cgs units. Also plotted in each quadrant are 
contour levels of the magnetic field amplitude in Gauss units.
Notice the field contours track the shock structure as it evolves, but the magnetic pressure does not 
amplify significantly enough to disrupt reactive flows nor affect burn product distributions in any significant way.

\begin{figure}
\hspace{1.5in}\includegraphics[width=0.5\textwidth]{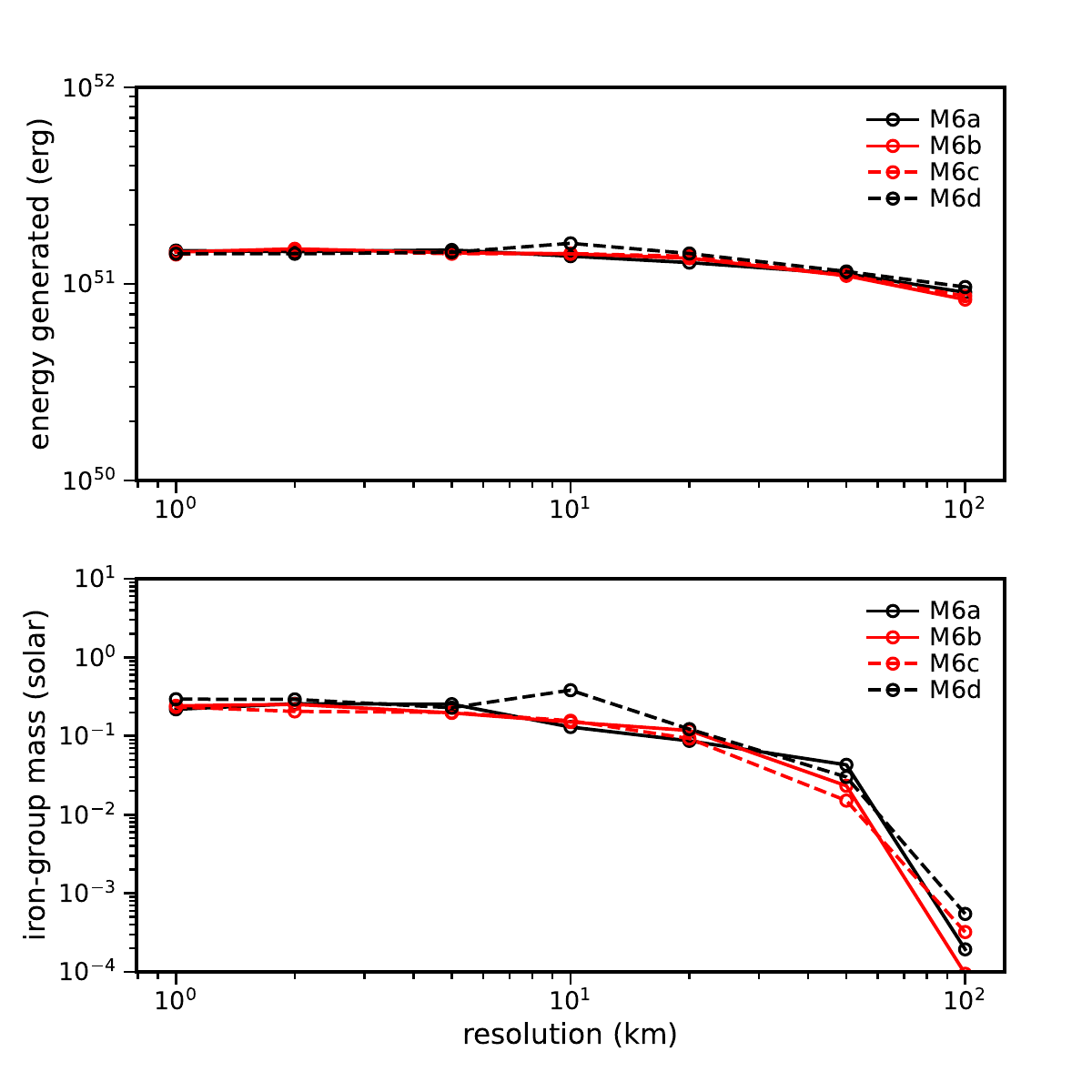}
\caption{
As Figure \ref{fig:wdwdres} except here we compare collisions of 0.6 $M_\odot$ 
magnetized and unmagnetized stars, and from different initial separations. 
Black (red) lines correspond to contact (L1 Lagrange point) separations,
and solid (dashed) lines represent unmagnetized (magnetized) configurations.
}
\label{fig:wdwdresB}
\end{figure}

We additionally evaluate how or if tidal forces developing from greater initial separations affect
the near-contact results from the previous section by placing the  mass centers in models M6b and  M6c
just outside the L1 Lagrange point (corresponding to nearly four star radii).
The results are presented in Figure \ref{fig:wdwdresB} showing the energy released and iron-group masses
from four calculations: unmagnetized at contact (M6a, solid black line)), unmagnetized at L1 separation (M6b, solid red line),
magnetized with a poloidal field at contact (M6d, dashed black line), and
magnetized with a toroidal field at L1 (M6c, dashed red line).
Results from the highest resolution runs are tabulated in Table \ref{tab:results_2d},
demonstrating excellent agreement with the models in
the previous section despite differences in separations and the inclusion or configuration of magnetic fields.

\subsubsection{Ignition Sites}
\label{subsubsec:ignitionSites}

\begin{figure}
\vspace{-0.0in}
\hspace{+0.5in}\includegraphics[width=1.0\textwidth]{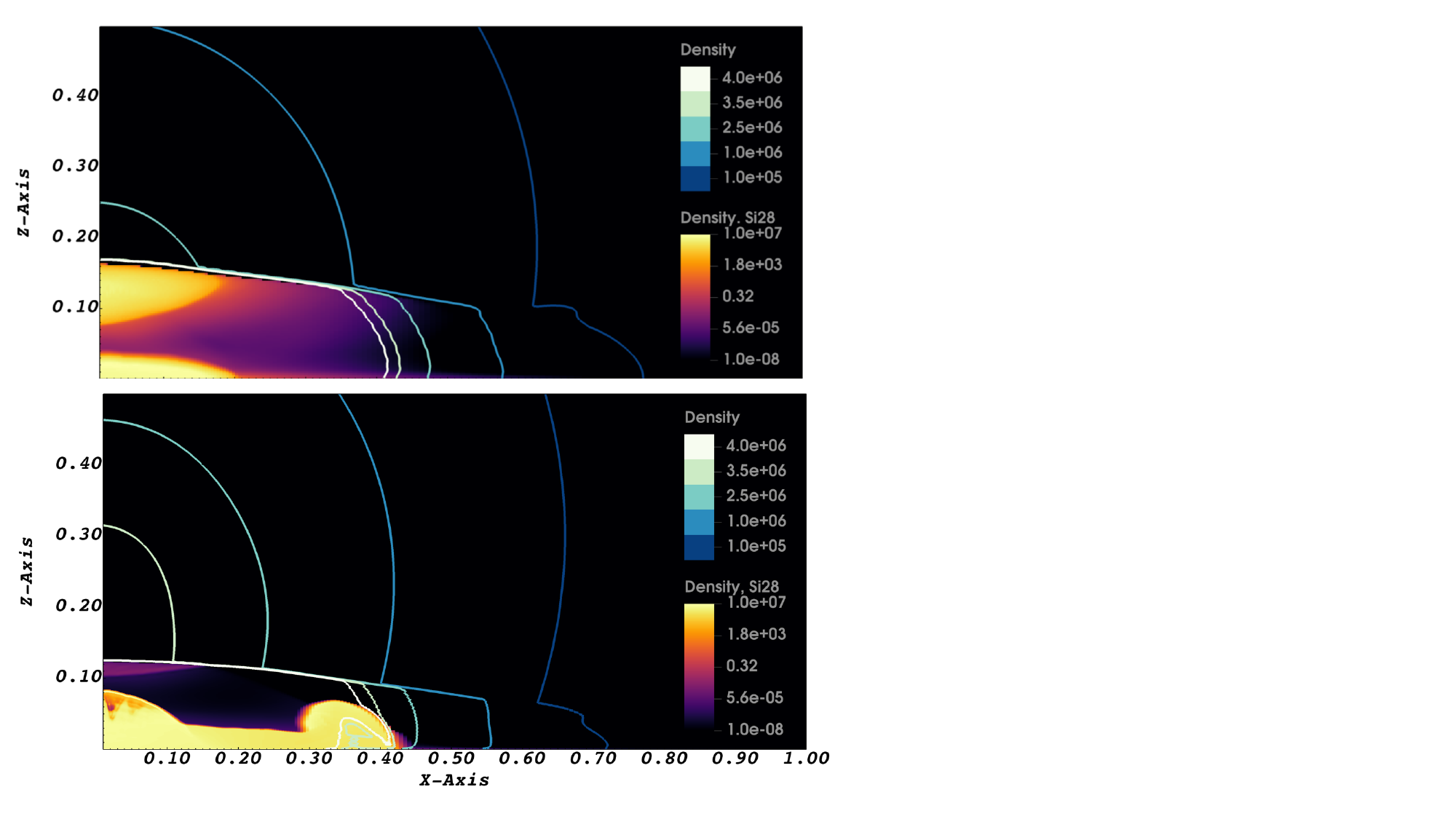}
\vspace{-0.0in}\caption{
The effect of grid resolution on the development, orientation, and timing of multiple, nearly simultaneous,
ignition sites is demonstrated by imaging the silicon density from two calculations differing only by resolution:
20 km (top, shown at 2.9 seconds) and 2 km (bottom, at 2.4 seconds). This particular image corresponds to
the poloidal magnetized model M6d, but is typical of unmagnetized models as well.
The five contours represent total density in cgs units, including all elements.
As usual, axes labels are in units of $10^4$ km.
}
\label{fig:plate_ignition}
\end{figure}

The relative smoothness of the convergence plots can mask
important differences in the development, orientation, timing, and number of ignition sites.
For example, our calculations find two distinctly different evolutionary tracks distinguishing 
low ($>5$ km) from high ($<5$ km) grid resolution. All calculations appear to develop multiple
shock-triggered ignition sites: a primary site at the origin or initial point of contact, and
delayed secondary sites further out in radius. However, as demonstrated in Figure \ref{fig:plate_ignition},
the location of these secondary sites depends on resolution, forming a few thousand
kilometers from the primary ignition site either along the pole axis at coarse resolution, 
or along the reflection plane at high resolution. 

Both plates of Figure \ref{fig:plate_ignition}
image the density of silicon from the M6d series as a diagnostic of early ignition, along with contours of the total
gas density demonstrating the compression state of the star. The top plate corresponds to 20 km resolution,
the bottom to 2 km. These images are plotted on the same spatial scale, zoomed into the interior of the star to
identify the shock structure, but are taken at two different times separated by approximately 0.5 seconds
(2.9 secs in the top plate, 2.4 secs in the bottom). Both images exhibit a common initiation site at the
origin, and secondary sites offset from the origin along orthogonal axes. In both cases, the secondary
sites are triggered by shock compression and heating: the coarse resolution case is triggered by
the upward moving shock generated at the point of contact, and the fine resolution case
is triggered by the transverse moving shock while compressing and ejecting material outward along the equatorial plane.
The total gas density contours indicate the approximate location of the enveloping (multi-dimensional) shock structure.
At coarse resolution, compression heating along the transverse direction ($x$-axis) is not adequately captured
so the star does not develop secondary hot spots until later in time when the upward moving shock
and the gravitationally collapsing star conspire to produce shock heated densities and temperatures
high enough to initiate silicon burn along the pole. 

\begin{figure}
\hspace{1.5in}\includegraphics[width=0.5\textwidth]{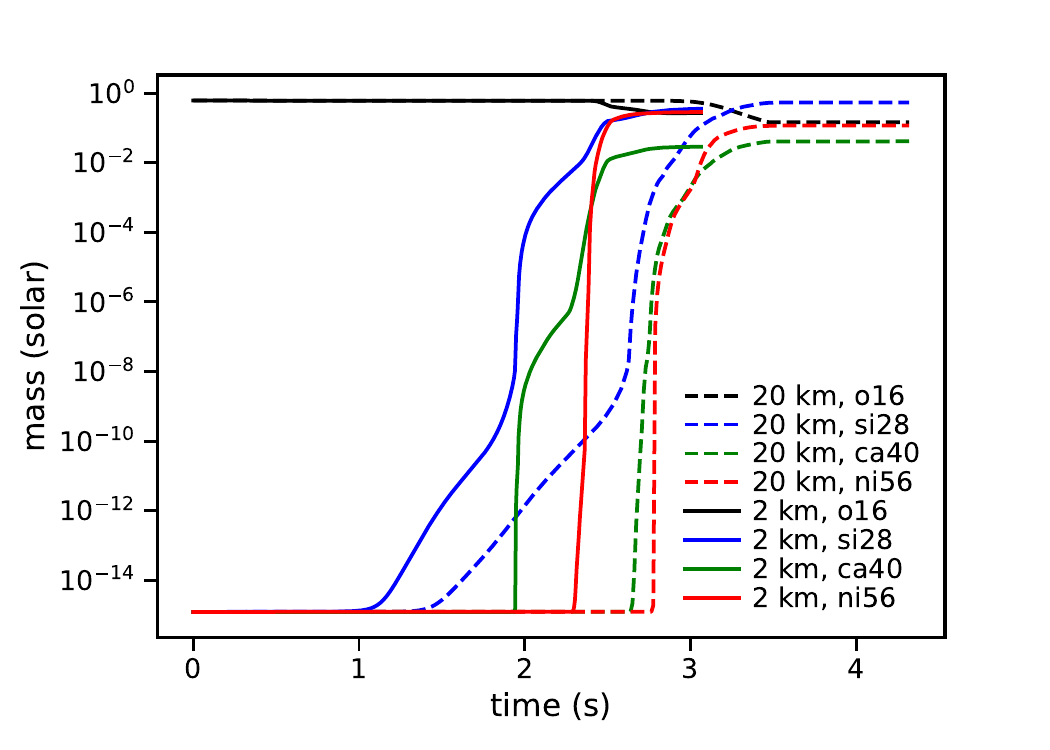}
\caption{
Total mass of a few select elements from model M6d (oxygen in black,
silicon in blue, calcium in green, and nickel in red) as a function of time.
Solid and dashed lines represent results from the 2 and 20 km resolution
calculations from Figure \ref{fig:plate_ignition}, correlating
ignition time to grid resolution.
}
\label{fig:wdwd_species}
\end{figure}

The delay in ignition time (roughly 0.5 seconds) due to grid resolution is shown in Figure \ref{fig:wdwd_species} where
we plot the time evolution of the total masses of oxygen, silicon, calcium, and nickel
for the two runs at 20 km resolution (dotted lines) and 2 km (solid lines). Despite
these starkly different evolutionary tracks (in both timing and the development of hot spots), the
end results, quantified by the total nuclear energy released and production yields, appear
not to be especially sensitive to these differences. 
\replaced{
We attribute this to the fact that once
ignition is acheived, regardless of where and when, there is sufficient C/O fuel reserves
pre-conditioned by gravitational free-fall to sustain thermonuclear burn throughout the star interior.
} {
We attribute this to the relative
proximity of the multiple nucleating sites to each other and to the point of contact, allowing
the different detonation sources to quickly fuse into a single, outwardly propagating burn front.
There are sufficient C/O fuel reserves pre-conditioned by gravitational free-fall
to sustain thermonuclear burn throughout the entire star
once ignition is acheived, regardless of the initiation sequence.
}

\subsubsection{Mesa Compositions and Helium Shells}
\label{subsubsec:mesa}

\begin{figure}
\hspace{1.5in}\includegraphics[width=0.5\textwidth]{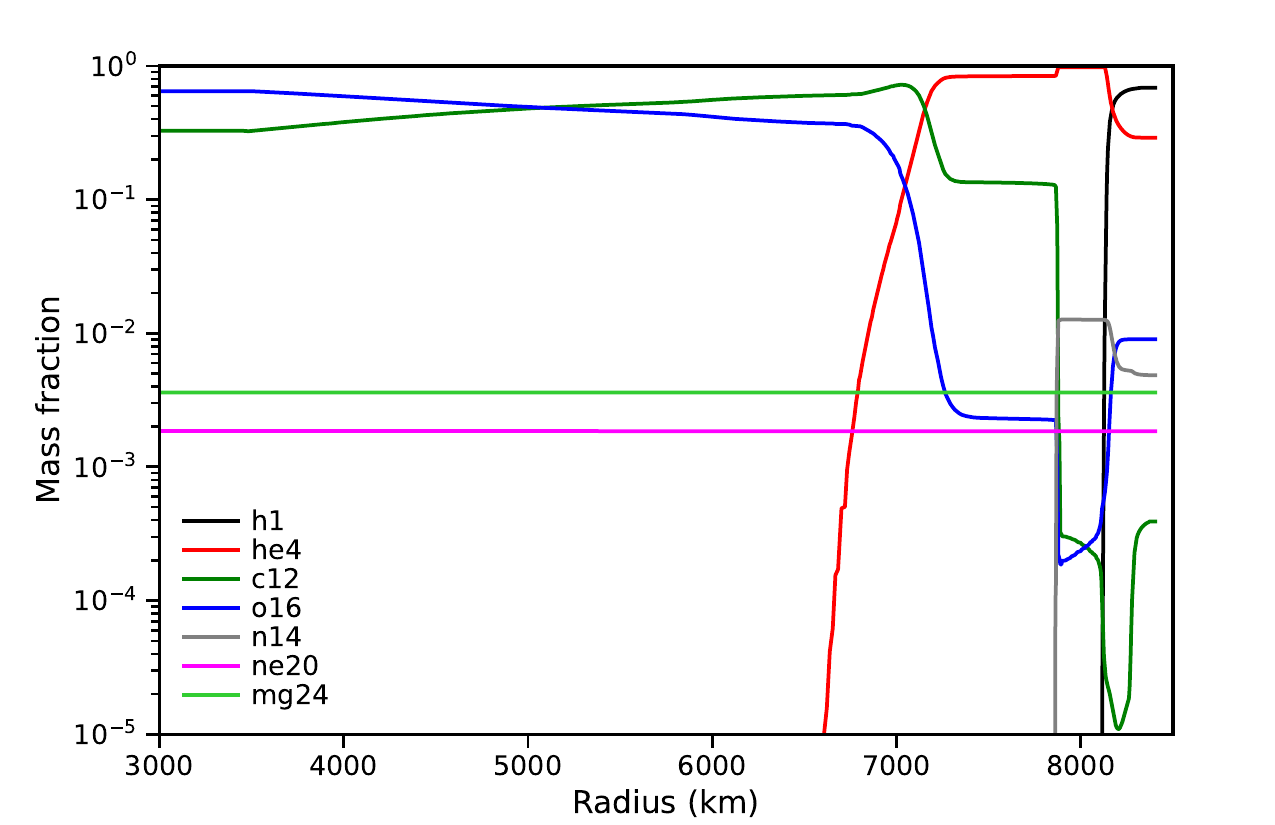}
\caption{
Dominant composition profiles derived from the Mesa stellar evolution code for a 0.6 solar mass WD as a function of radius.
}
\label{fig:wdwd_initialSpecies}
\end{figure}

Because of the high uncertainty exhibited by the 1D reactive He/C/O shocktube tests in section \ref{subsec:burntube},
we explore here the effect of introducing helium to our initial WD model compositions. We consider two
compositions: the first model M6e is derived directly from the Mesa stellar evolution code \citep{Paxton11},
\added{
mapped onto the grid as described in \cite{Anninos19},
}
and plotted in Figure \ref{fig:wdwd_initialSpecies} for a 0.6 $M_\odot$ WD; the second M6g utilizes a 50/50 mixture of C/O in the
interior of the star but replaces all elements in the outer 2000 km shell with pure helium.
We anticipate (and observe) model M6g to undergo double detonation ignition,
burning simultaneously radially outward from the C/O core as well as spherically through the 
helium shell \citep{Townsley19,Boos21}.
\added{
Both models were calculated without and with the artificial filter near the equatorial plane (which
added stability to the 1D shocktube tests). Interestingly we find the filter has no significant effect
on convergence or final production outcomes. What was unpredictable or random in the shock tube test was the
location and timing of the initiating hotspot. However, and as we pointed out in the previous section,
once detonation triggers in these models it matters little how far off the equatorial plane it first ignited.
}

\begin{figure}
\vspace{-0.0in}
\hspace{+0.2in}\includegraphics[width=0.8\textwidth]{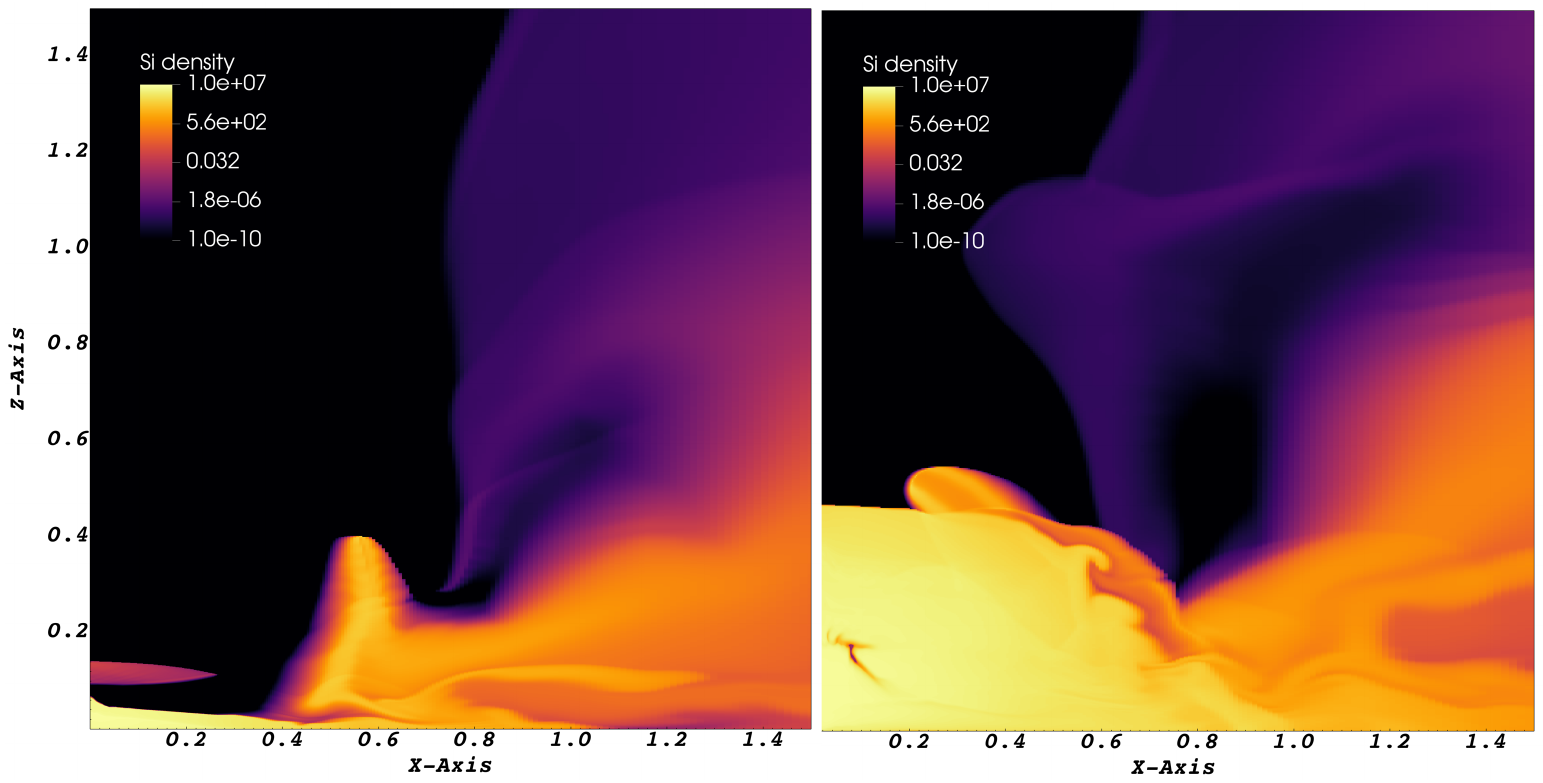}
\vspace{-0.0in}\caption{
Silicon density from model M6g imaged at two different times: 
2.6 seconds (left) and 3.4 (right). The early image shows clear evidence of multiple and nearly simultaneous
initiation sites, including the stellar center, the vertical offset immediately behind the upward moving
hydrodynamic shock, and the detonation running through the outer helium shell. The late image shows the
helium shell front reaching the opposite axis at about the same time as the 
main (stronger) detonation traveling radially outward from the star center.
}
\label{fig:plate_heshell}
\end{figure}

Figure \ref{fig:plate_heshell} shows a couple of time instances from the double detonation model M6g.
Two prominent burn fronts ignite nearly simultaneously and propagate separately through the helium shell 
along the outer surface of the star and radially outward from the center of the C/O core. This figure plots the silicon
density corresponding to the 2 km resolution case, but we note that the 1 km case behaves identically:
The two wave fronts burn robustly enough to reach the opposite ends of the star at about the same time.
However, the robustness and sustainability of the detonation front across the outer shell depends strongly on grid resolution.
For example, at intermediate resolutions (5 to 10 km) the surface wave front propagates slower than
in the high resolution (1 to 2 km) cases, and is swept up
by the stronger (core) detonation very early. At even coarser resolutions ($\ge 20$ km), hotspots are created
in the helium shell near the collision plane, but the burning cannot sustain itself and the front essentially freezes
in place (does not propagate) until the core detonation overtakes it. Additionally we find that even for
the highest resolution models, the helium shell does not produce significant amounts of iron group products
but is nevertheless a strong source of intermediate mass elements, from silicon to calcium.

\begin{figure}
\hspace{1.5in}\includegraphics[width=0.5\textwidth]{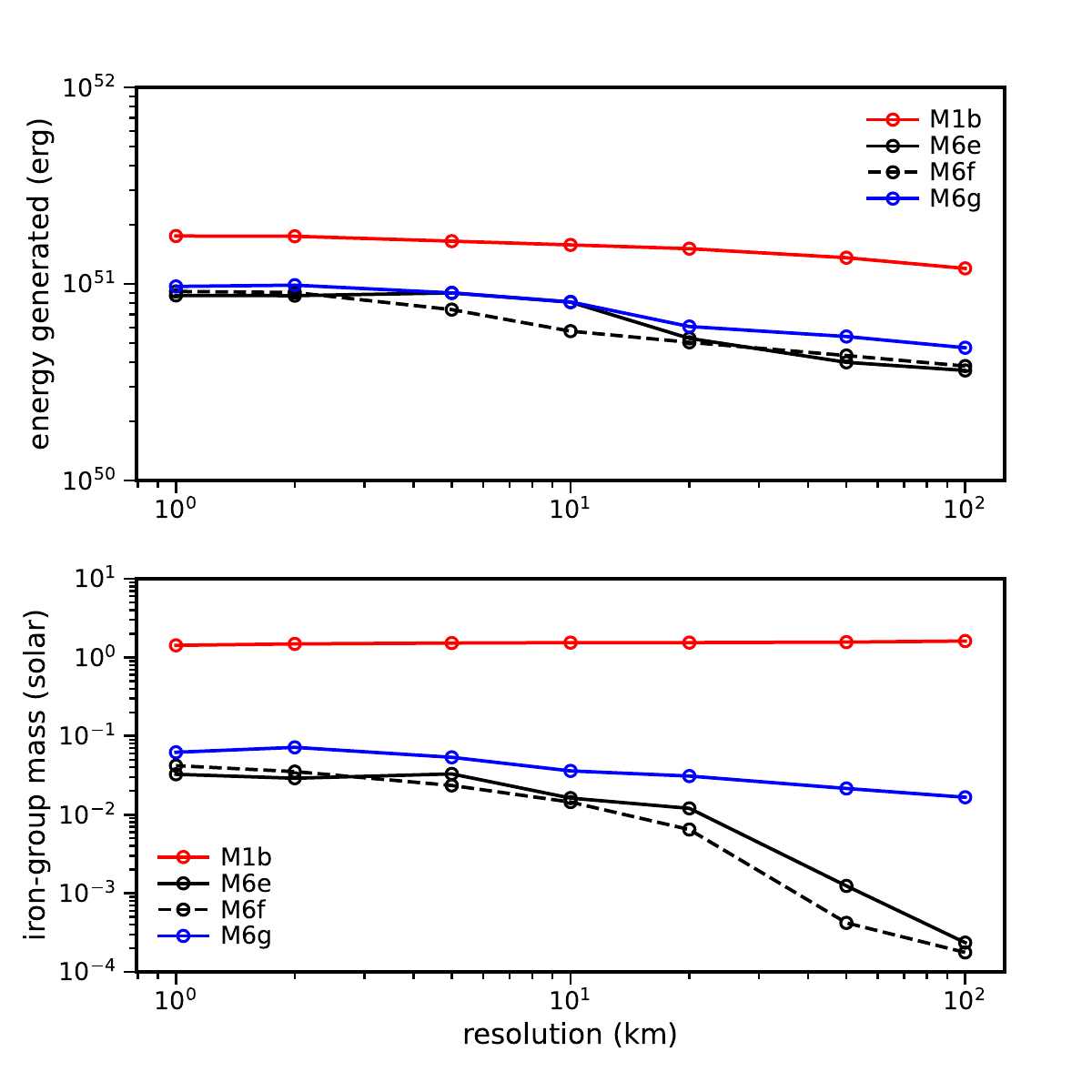}
\caption{
As Figures \ref{fig:wdwdres} and \ref{fig:wdwdresB} except for the He/C/O composition cases M1b (Mesa composition),  
M6e (Mesa composition), M6f (40/60 C/O), and M6g (50/50 C/O with He shell).
}
\label{fig:wdwdresmesa}
\end{figure}

Figure \ref{fig:wdwdresmesa} plots the total nuclear energy released (top) and iron-group production (bottom)
comparing models M1b, M6e, M6f, and M6g as a function of grid resolution. We find convergence patterns similar
to the pure C/O models, except at relatively diminished values with fewer burn products and less released energy.
However, differences in the conversion efficiency are attributed to different C/O mixtures not to helium content
as observed by comparing the excellent agreement between models M6e and M6f, 
a pure C/O WD composed of roughly the same 40/60 mixture as M6e.
This behavior is typical of carbon-depleted cores as they generally require higher imploding shock strengths
to sustain detonations (\cite{Seitenzahl09} for example finds larger critical hotspot radii in reduced carbon mixtures).

\section{Conclusions}
\label{sec:conclusions}

We have explored the stability, reliablity, and general convergence
behavior of nuclear reactive flow calculations from the head-on collision of two equal mass white dwarf stars.
Our motivation is derived from calculations in the literature that have not always
agreed quantitatively on important diagnostics like the amount of iron-group products produced in these
scenarios. We are also somewhat skeptical about claims of convergence, and the always-worrisome possibility that the
application of limiters might have unintended consequences, particularly if they are applied arbitrarily
and without careful study.

For this work we considered equal mass collisions of 0.6 $M_\odot$ and 1 $M_\odot$ stars, simulated on geometrically
refined grids specially designed to concentrate resources at initiation sites along the equatorial and polar planes.
Calculations were carried out at resolutions ranging from 100 to 1 km, and stabilized with several
reaction limiting procedures, including by energy released, temperature variations, and local reactant changes.
We find generally consistent results independent of limiter procedures (or what field is limited)
when limiter amplitudes are bounded on both the low and high ends. For example,
at $f_E < 0.01$ the hydrodynamic solutions maintain their convergent properties but the strong
limiting of energy creation begins to affect the thermodynamic behavior, resulting in solutions deviating
more and more from the temperature-limited solutions. Additionally, although energy limiting tolerates
large amplitudes, limiting by temperature does not, so that values of $f_T > 0.05$ tend to produce
unstable or at best unreliable solutions. 

Across the restricted range $0.01 < (f_T, ~f_E) < 0.05$, we found all versions of limiting
procedures produced consistently stable solutions, with well-behaved, monotonic trends 
towards convergence with uncertainties better than 10\% at 5 km resolution
(when compared to the 1 km results) for total energy
released and iron-group mass, and about 20\% for intermediate (calcium-group) elements.
Increasing resolution stabilizes the number, location, and development of initiation sites,
as well as securing the sustainability of detonation waves through helium surface shells.
These results hold for both unmagnetized and magnetized models, for simple C/O or more complex
Mesa core compositions, with or without outer helium shells,
and regardless of whether the two stars are initialized at contact or separated by multiple stellar radii.

Similar conclusions are drawn from our 1D reactive shock tests.
There we also found evidence of clear convergence at a few kilometers resolution,
and inferred comparable (lower and  upper bound) restraints on limiter amplitudes.
We point out that these results are generally consistent with earlier work on axisymmetric
collisions of WDs \citep{Kushnir13}, and with tidal disruptions of stars by black holes \citep{Anninos22}.
The latter applied a similar analysis approach to a very different binary interaction, but
arrived at essentially the same conclusions.

\begin{acknowledgments}

This work was performed under the auspices of the U.S. Department of Energy by 
Lawrence Livermore National Laboratory under Contract DE-AC52-07NA27344. 

\end{acknowledgments}

\software{ Cosmos++ \citep{Anninos05,Anninos17,Anninos20,Roth22}, Torch \citep{Timmes99} }

\bibliography{refs_wdwd}{}
\bibliographystyle{aasjournal}

\listofchanges

\end{document}